\documentclass[fleqn,usenatbib]{mnras}

\usepackage{newtxtext,newtxmath}

\usepackage[T1]{fontenc}
\usepackage{ae,aecompl}

\usepackage[utf8]{inputenc}
\usepackage{graphicx}	
\usepackage{amsmath}	

\newcommand{\fermi}{\textit{Fermi}-LAT}

\title[Blazar jets with similar baryon loading]{Blazar jets launched with similar energy per baryon, independently of their power}

\author[J.M. Rueda-Becerril, A.O. Harrison and D. Giannios]{Jesús M. Rueda-Becerril$^{1}$\thanks{E-mail: jruedabe@purdue.edu}, Amanda O. Harrison$^{1}$, Dimitrios Giannios$^{1}$
\\
$^{1}$Department of Physics, Purdue University, 525 Northwestern Avenue, West Lafayette, IN, 47907, USA\\
}

\date{Accepted XXX. Received YYY; in original form ZZZ}

\pubyear{2015}

\begin{document}
\label{firstpage}
\pagerange{\pageref{firstpage}--\pageref{lastpage}}
\maketitle

\begin{abstract}
The most extreme active galactic nuclei (AGN) are the radio active ones whose relativistic jet propagates close to our line of sight. These objects were first classified according to their emission line features into flat-spectrum radio quasars (FSRQs) and BL Lacertae objects (BL Lacs). More recently, observations revealed a trend between these objects known as the \emph{blazar sequence}, along with an anti-correlation between the observed power and the frequency of the synchrotron peak. In the present work, we propose a fairly simple idea that could account for the whole blazar population: all jets are launched with similar energy per baryon, independently of their power. In the case of FSRQs, the most powerful jets, manage to accelerate to high bulk Lorentz factors, as observed in the radio. As a result, they have a rather modest magnetization in the emission region, resulting in magnetic reconnection injecting a steep particle energy distribution and, consequently, steep emission spectra in the $\gamma$-rays. For the weaker jets, namely BL Lacs, the opposite holds true; i.e., the jet does not achieve a very high bulk Lorentz factor, leading to more magnetic energy available for non-thermal particle acceleration, and harder emission spectra at frequencies $\gtrsim$ GeV. In this scenario, we recover all observable properties of blazars with our simulations, including the \emph{blazar sequence} for models with mild baryon loading ($50 \lesssim \mu \lesssim 80$). This interpretation of the blazar population, therefore, tightly constrains the energy per baryon of blazar jets regardless of their accretion rate.
\end{abstract}


\section{Introduction}
\label{sec:intro}

Blazars are a subclass of radio-loud AGNs with a relativistic jet propagating close to the line of sight of the observer. The emission from these objects covers all frequencies of the electromagnetic spectrum, producing a double bump  structure. The peak of the low frequency bump ranges from infrared to X-ray, whereas the high frequency one peaks in the $\gamma$-ray. Blazars have been classified into two subclasses based on the properties of their emission lines: FSRQs and BL Lacs \citep{Urry:1995pa}. Blazar science has greatly advanced, during the last decade, thanks to dedicated monitoring programs at different wavelengths \citep[e.g.,][]{Ghisellini:2010ta,Blinov:2015pa,Lister:2016mo,Jorstad:2016ma,Ackermann2011,Rani:2017st}. In part because \fermi\ allowed, for the first time, for the systematic study of the populations as a whole by following an unprecedented number of sources in $\gamma$-rays \citep{Ackermann2011,Ajello:2014ro}. Therefore, we can now move beyond the case-by-case studies and attempt a holistic approach in understanding the physical processes involved. One of the clear trends identified by \fermi\ is that BL Lac objects are characterized, on average, by harder spectra than FSRQs \citep{Ghisellini:2009ma}. As a result, BL Lac objects are the most extreme TeV emitters \citep{Ajello:2014ro}. BL Lacs are also typically characterized by a synchrotron peak at higher energies (as high as X-rays). Not surprisingly, modeling of the spectrum of blazars requires electrons injected with much higher energies in BL Lacs than in FSRQs \citep{Celotti:2008gi}.

The systematic differences of the two blazar classes are not limited to their $\gamma$-ray properties. Radio programs like MOJAVE have shown that FSRQs are characterized by extreme apparent speeds ($\beta_{\rm app} \sim$ 10) in contrast to those of BL Lacs ($\beta_{\rm app} \sim$ 2) \citep{Kovalev:2009al,Homan:2009ka,Lister:2009co,Lister:2011al,Lister:2019ho}. Also, BL Lacs are likely associated with less powerful jets (FR-I equivalent) in contrast to FSRQs (FR II equivalent) \citep{Ghisellini:2001ce,Giommi:2012pa,Giommi:2013pa,Giustini:2019pr}. It has also been pointed out that the luminosity of the broad-line region (BLR) may be a distinctive between the two kinds of blazars \citep[e.g.,][]{Ghisellini:2001ce,Ghisellini:2009ma,Ghisellini:2011ta,Ghisellini:2008ta}, as well with other intrinsic parameters such as the spin of the black hole \citep{Mier:2002na,Tchekhovskoy:2010na,Garofalo:2019re}. A main parameter in these models is the accretion rate $\dot{M}$ onto the black hole. Let us introduce here the Eddington rate
\begin{equation}\label{eq:dotm}
  \dot{m} \equiv \dfrac{\dot{M}}{\dot{M}_{\rm Edd}},
\end{equation}
where $\dot{M}_{\rm Edd}$ is the Eddington mass accretion rate (see Sec.~\ref{sec:AccrJetLumin}). Therefore, $\dot{m}$ gives a measure of the accretion rate of the AGN as a fraction of the Eddington rate. In this work we will use $\dot{m}$ to differentiate BL Lacs from FSRQs, so that BL Lac objects would be those blazars with low $\dot{m}$, while FSRQs those with high $\dot{m}$.

The so called \emph{blazar sequence} \citep[][]{Padovani:2007ap} has been of strong observational and theoretical focus since the first multiwavelenght spectral energy distributions (SEDs) of different objects were compared \citep{Fossati:1998ma,Ghisellini:1998ce}. Evolutionary scenarios have been proposed in the past decades which connect both kinds of objects in terms of accretion efficiency and the jet formation \citep{Bottcher:2002de,Maraschi:2003ta,Celotti:2008gi,Ghisellini:2011ta}. Thanks to \fermi\ observations the view of the \emph{blazar sequence} has evolved and more sophisticated trends have been proposed since its introduction \citep[e.g.][]{Meyer:2011fo,Finke:2013co,Ajello:2014ro,RuedaBecerril:2014mi}. Furthermore, recent works have questioned if those trends correspond to continuum transition between the two kinds of blazars \citep{Padovani:2019oi,Keenan:2020me}.

On the theoretical front, AGN jets are believed to be launched magnetically dominated in the vicinity of a rotating black hole \citep{Blandford:1977zn}. Magnetohydrodynamic (MHD) simulations of jet acceleration predict that the bulk acceleration of the jet takes place at the expense of its magnetization, i.e., while the bulk Lorentz factor $\Gamma$ of the jet increases, its magnetization $\sigma$ (defined as the Poynting flux to the total energy flux ratio of the jet) decreases \citep[][]{Komissarov:2007ba,Komissarov:2009vl,Tchekhovskoy:2008mc}. According to observations, FSRQs appear with a bulk Lorentz factor $\Gamma$ of a few tens, in contrast to the slower BL Lacs (see, e.g., \cite{Homan:2009ka}). This means that FSRQs appear to be associated with more efficiently accelerated jets, leaving a low energy budget per baryon in the emission region. This is in contrast to BL Lacs which do not reach as large of a bulk Lorenz factor but, as a result, have an emission region of high magnetization.

It is clear from observations that AGN jets may propagate as far as several kpc to a few Mpc from the central engine. Relativistic hydrodynamic and MHD simulations have shown that it is highly probable that instabilities may develop in relativistic jets \citep{Perucho:2006lo,LopezCamara:2013mo,Matsumoto:2013ma,Komissarov:2019go,Tchekhovskoy:2016br}. Instabilities may translate into dissipation of energy. In particular, if kink instabilities develop in the jet, this could translate into a tangled magnetic field in the jet \citep{Tchekhovskoy:2016br,BarniolDuran:2017tc}. This could in turn induce the formation of current sheets, allowing to trigger magnetic reconnection. The theory of magnetic reconnection in the context of blazar flares has been explored in the past several years \citep{Giannios:2009uz,Nalewajko:2011gi,Sironi:2015pe,Petropoulou:2016gi,Christie:2019pe}, showing that it may be the process responsible for the non-thermal particle acceleration and radiation \citep{Spruit:2001da,Giannios:2006sp,Sironi:2014sp,BarniolDuran:2017tc}. In recent years, first-principle particle in cell (PIC) simulations have demonstrated that magnetic reconnection can account for many of the extreme spectral and temporal properties of blazars \citep{Sironi:2014sp,Sironi:2015pe,Petropoulou:2016gi,Christie:2019pe}. Interestingly, these simulations have shown that the crucial parameter that controls the distribution of accelerated particles is the jet magnetization $\sigma$. Even for a modest increase in $\sigma$ of the plasma, magnetic reconnection results in much harder particle distributions, and, as a result, harder emission spectra \citep{Petropoulou:2016gi,Petropoulou:2019si}.

In this work we will not focus on the details of the structures that form in the current sheet but only on the global properties of the emission region. To determine the fraction of magnetic energy that is dissipated in the reconnection region and the resulting particle distributions, we will exploit the findings of \citet{Sironi:2015pe} and subsequent work. These studies provide specific predictions for the distribution of the accelerated particles as a function of the jet magnetization $\sigma$. The clear trend is that for $\sigma \lesssim 10$, the resulting particle spectra are described by a steep power-law distribution function $\gamma'^{-p}$, where the slope $p \gtrsim 2$. A soft particle energy distribution results in low energy peaks for characteristic emission bumps as well as softer resulting spectra. This scenario would correspond to FSRQs which, as we have mentioned before, have a modest magnetization at the emission region. On the other hand, a strongly magnetized jet such as a BL Lac ($\sigma\gtrsim 10$) would be characterized by a hard spectrum of accelerated particles with $1 \lesssim p \lesssim 2$.

The setup of our model is described in Sec.~\ref{sec:model}, along with its most relevant parameters, and a brief description of the numerical code employed. In Sec.~\ref{sec:results} we present and describe the results obtained out of our simulations. Finally, in Sec.~\ref{sec:discuss} we discuss the model, the results, its implications, and in Sec.~\ref{sec:conclusions} we make the final conclusions from this study.

\section{Model}
\label{sec:model}

According to MHD theory of relativistic jets, a quantity which is conserved along magnetic  field lines is the total energy flux per unit rest-mass energy flux $\mu$ \citep[see][]{Komissarov:2007ba,Tchekhovskoy:2009mc}, also known as the baryon loading. For a cold plasma flow:
\begin{equation}\label{eq:mu}
  \mu = \Gamma (1 + \sigma),
\end{equation}
where $\Gamma$ and $\sigma$ are the flow bulk Lorentz factor and magnetization, respectively. The magnetization $\sigma$ is defined as the ratio between the Poynting flux and the hydrodynamic energy flux.
\begin{equation}
  \sigma = \dfrac{B'^{2}}{4 \piup \rho' c^2},
\end{equation}
where $B'$ and $\rho'$ are the magnetic field strength and the mass density of the plasma\footnote{Quantities measured in the comoving frame of the fluid will be denoted with a prime sign ('), unless noted otherwise. Quantities measured by a cosmologically distant observer will be denoted with the subscript `obs'. Quantities measured in the laboratory frame will remain unprimed.}.

In this section we will describe a simple model from which we are capable of accounting for the \emph{blazar sequence} by just considering a simple relation between the jet power and bulk Lorentz factor $\Gamma$, where more powerful jets are the fastest. We assume that both the jet luminosity $L_{\rm j}$ and the bulk Lorentz factor $\Gamma$ depend only on the accretion rate parameter $\dot{m}$, keeping the baryon loading $\mu$ as a free parameter. This setup strongly constrains/binds the magnetic and kinetic properties of the emission region. We will quantitatively test this picture and show that the blazar sequence can be simply understood in a scenario where $\mu$ changes little among different objects.

\subsection{Accretion and jet luminosities}
\label{sec:AccrJetLumin}

Let us define the radiative efficiency of the disk $\eta_{\rm d} \equiv L_{\rm d} / \dot{M} c^{2}$ \citep[e.g.,][]{Davis:2011la}, where $c$ is the speed of light, and $L_{\rm d}$ the disc luminosity. From this parameter let us define the Eddington mass accretion rate as follows:
\begin{equation}\label{eq:dotM-Edd}
  \dot{M}_{\rm Edd} \equiv \dfrac{L_{\rm Edd}}{\eta_{\rm d} c^{2}},
\end{equation}
where $L_{\rm Edd} \approx 1.26 \times 10^{36} (M / M_{\odot})$~erg~s$^{-1}$.
The jet luminosity $L_{\rm j}$ is related to the accretion power by \citep[e.g.,][]{Celotti:2008gi}
\begin{equation}\label{eq:Lj-dotM}
  L_{\rm j} = \eta_{\rm j} \dot{M} c^{2}
\end{equation}
where $\eta_{\rm j}$ is the jet production efficiency. From equations \eqref{eq:Lj-dotM} and \eqref{eq:dotM-Edd} we get that
\begin{equation}\label{eq:Mdot-Medd-rat}
  L_{\rm j} = \dfrac{\eta_{\rm j}}{\eta_{\rm d}} L_{\rm Edd} \dot{m}.
\end{equation}

According to radio observations there seems to be a correlation between the bulk Lorentz factor of the emission region and the jet power \citep{Lister:2009co,Homan:2009ka}, or $\dot{m}$ for this effect, according to Eq.~\eqref{eq:Mdot-Medd-rat}. Out of these empirical relation we make the following ansatz:
\begin{equation}\label{eq:dotm-gamma}
  \dot{m} = {\left( \dfrac{\Gamma}{\Gamma_{0}} \right)}^{s}.
\end{equation}
It is worth noting here that the parameter $\Gamma_0$ has no particular physical meaning. This parameter results from the proportionality relation between $\dot{m}$ and $\Gamma$. In other words, the bulk Lorentz factor of the jet is regulated by the Eddington ratio. In this study we assume that accreting black holes in AGNs are at most Eddington luminous. From observations \citep[e.g.][]{Lister:2019ho} we therefore set $\Gamma_0 = 40$. In order to estimate the values of $s$ we performed a series of simulations varying $s$ between 1.5 and 4.0. We find that the simulation outcomes do not vary significantly for $2.5 \lesssim s \lesssim 3.5 $ (for further details see Appendix~\ref{sec:acc-rate}). Hence we set $s = 3.0$, which gives
\begin{equation}\label{eq:dotm-num}
  \dot{m} \approx 1.56 \times 10^{-5}\, \Gamma^3.
\end{equation}

\subsection{External radiation field}
\label{sec:ext-rad}

According to the standard model of AGNs \citep{Urry:1995pa}, the material pumped into the jet will often move through an external radiation field produced by the Broad Line Region (BLR). The BLR is believed to be reprocessed radiation from the accretion disk \citep{Sikora:1997ma,Tavecchio:2008gh}. The radius, size and geometry of the BLR are still a topic of debate, although it has been thoroughly studied over the last decades \citep[e.g.,][and references therein]{Kaspi:2005ma,Kaspi:2007br,Gaskell:2009rv}. As mentioned above, BL Lacs are considered to have a low-Eddington accreting black hole, which translates into a faint BLR radiation field; opposed to FSRQs whose black hole is considered to be accreting at higher rates, and therefore a larger density of reprocessed photons in the BLR.

The precise localization of the emission region is still under debate. Different models locate the dissipation either below the BLR \citep{Tavecchio:2008gh} or outside the BLR \citep{Marscher:1985ge}. BL Lacs, for instance, may easily be accounted for with the latter. Whereas FSRQs may not, since in outer regions there will be less photons to be upscattered through IC. In the present study we will assume that energy dissipation takes place within the BLR \citep[e.g.,][]{Sikora:1997ma,Georganopoulos:2005ka}. In our model we will assume that the emission region is immersed in an isotropic and monochromatic radiation field. The energy density of the external BLR radiation can be parametrized as follows \citep{Ghisellini:2008ta}:
\begin{equation}\label{eq:uBLR}
  u_{\rm BLR} = \eta_{\rm BLR} \dfrac{L_{\rm d}}{4 \uppi c R_{\rm BLR}^2}
\end{equation}
where $R_{\rm BLR} \simeq 10^{17} L_{{\rm d},45}^{1 / 2}$~cm is the radius of the BLR, $\eta_{\rm BLR}$ the covering factor, and $L_{{\rm d}, 45} = L_{\rm d} / (10^{45} \, \mathrm{erg \, s^{-1}})$. Finally, we will consider the radiation field in this region to be monochromatic with frequency $\nu_{\rm BLR}$. In the comoving frame of the plasma flow, $\nu'_{\rm BLR} = \Gamma \nu_{\rm BLR}$ and $u'_{\rm BLR} = \Gamma^2 (1 + \beta^2 / 3) u_{\rm BLR}$, where $\beta \equiv \sqrt{1 - \Gamma^{-2}}$ is the bulk speed of the flow in units of the speed of light.

\subsection{On the jet composition and emission region}
\label{sec:EmisReg}

Let us consider an electron-proton jet. According to MHD theory, instabilities in a Poynting flux dominated flow (i.e., with $\sigma \gtrsim 1$) lead to the formation of current sheets, where magnetic reconnection is triggered \citep[see][]{Eichler:1993ma,Begelman:1998in,Giannios:2006sp}. In the last decade great progress has been made on the understanding of relativistic reconnection trough PIC simulations \citep{Sironi:2014sp,Sironi:2015pe,Petropoulou:2016gi}, showing that instabilities develop magnetic islands (plasmoids) in which particles accelerate to ultra-high energies due to magnetic energy dissipation \citep[see][for a review]{Kagan:2015si}.

The magnetization of a relativistic jet is defined as the ratio of the magnetic energy flux to the matter energy flux \citep[e.g.,][]{Janiak:2015si}
\begin{equation}
  \sigma = \dfrac{L_{\rm B}}{L_{\rm kin}} = \dfrac{L_{\rm B}}{L_{\rm j} - L_{\rm B}}.
\end{equation}
By solving the above equation for the Poynting flux luminosity we get that
\begin{equation}\label{eq:LB-sigma-Lj}
  L_{\rm B} = \dfrac{\sigma}{1 + \sigma} L_{\rm j},
\end{equation}
which in turn we use to calculate the magnetic energy density of the emitting blob in the comoving frame:
\begin{equation}\label{eq:uB}
  u'_{\rm B} = \dfrac{L_{\rm B}}{2 \uppi R_{\rm b}'^2 c \beta \Gamma^{2}},
\end{equation}
where $R_{\rm b}'$ is the size of the emission region or \emph{blazar zone}, assumed to be comparable to the cross section of the jet. We also assume that, over a dynamical time $t_{\rm dyn} \sim R'_{\rm b}/c$, a fraction $f_{\rm rec}$ of the magnetic energy in the blob is transferred to the electrons in the system in the form of kinetic energy. In other words, from Eq.~\eqref{eq:uB} we get that the luminosity of the electrons in the comoving frame of the blob reads:
\begin{equation}\label{eq:Le1}
	L_{\rm e}' = f_{\rm rec} \dfrac{2 L_{\rm B}}{3 \beta \Gamma^{2}}
\end{equation}

\subsubsection{The emission region}

In blazar jets, magnetic reconnection is believed to take place far from the central engine, but at sub-parsec scales \citep[e.g.,][]{Petropoulou:2016gi,Christie:2019pe}. We call such place the \emph{emission region}, which we will assume is at a distance $R_{\rm em}$ from the central engine, and to be a spherical blob in the comoving frame of the fluid, covering the cross-sectional area of the jet. We will also assume that the emission region is located close to the outer edge of the BLR, e.g., $R_{\rm em} = 0.9 R_{\rm BLR}$ \citep[see][]{Padovani:2019oi}. We can estimate the radius of the emitting blob, in the comoving frame of the flow, as follows:
\begin{equation}\label{eq:blob-radius}
  R_{\rm b}' \approx R_{\rm em} \theta_{\rm j},
\end{equation}
where $\theta_{\rm j} \approx 1 / \Gamma$ is the half-opening angle of the conical jet.

Let us take now a distant observer whose line of sight makes an angle $\theta_{\rm obs}$ with respect to the direction of motion of the emitting blob. Assuming that the blob emits isotropically \citep{Gould:1979cs}
\begin{equation}\label{eq:nuFnu}
  \nu L_{\nu} = \dfrac{3 f(\tau_{\nu'}')}{\tau_{\nu'}'} \mathcal{D}^{4} V' \nu' j_{\nu'}',
\end{equation}
where $\tau_{\nu'}' \equiv 2 R_{\rm b}' \kappa_{\nu'}$, $j_{\nu'}'$ and $\kappa_{\nu'}'$ are the synchrotron emissivity and self-absorption, respectively \citep{Rybicki:1979}, and
\begin{equation}
  f(\tau) \equiv \dfrac{1}{2} + \dfrac{\exp(-\tau)}{\tau} - \dfrac{1 - \exp(-\tau)}{\tau^2},
\end{equation}
is the optical depth function for a spherical blob \citep{Gould:1979cs,Dermer:2009}. The transformation from the comoving frame of the blob to the central engine reference frame is given by the Doppler factor: $\mathcal{D} \equiv {[\Gamma (1 - \beta \cos\theta_{\rm obs})]}^{-1}$.

\subsubsection{Particle acceleration}
\label{sec:PartAccel}

The magnetization of the plasma undergoing magnetic reconnection in the context of blazars has been studied thoroughly through PIC simulations in recent years \citep[e.g.][]{Sironi:2015pe,Sironi:2016gi,Petropoulou:2016gi}. As these simulations have shown, the energy distribution of accelerated electrons follows a power-law (non-thermal) profile:
\begin{equation}\label{eq:inj-term}
  Q'(\gamma') = Q_{0} \gamma'^{-p} H[\gamma'; \gamma'_{\min}, \gamma'_{\max}]
\end{equation}
where $\gamma'$ is the electrons Lorentz factor in the comoving frame, $H[x]$ the Heaviside function, and $\gamma'_{\min}$ and $\gamma'_{\max}$ are the minimum and maximum Lorentz factors of the distribution of accelerated electrons. The normalization factor $Q_{0}$ can be estimated by calculating the power of these electrons from Eq.~\eqref{eq:inj-term}, i.e.,
\begin{align}\label{eq:Le2}
  L'_{\rm e} & = V' Q_{0} m_{\rm e} c^{2}\int_{\gamma_{\min}}^{\gamma_{\max}'} {\rm d}\gamma \gamma^{-(p - 1)} \nonumber\\
  & = V' Q_{0} m_{\rm e} c^{2} \gamma_{\min}'^{2 - p} \mathcal{P}(\gamma_{\max}' / \gamma_{\min}', p - 1),
\end{align}
where $V' = (4 / 3) \uppi {R_{\rm b}'}^3$ is the volume of the emission region, and
\begin{equation}\label{eq:Pinteg}
  \mathcal{P}(a, s) := \int_{1}^{a} {\rm d}x\, x^{-s}
\end{equation}
is the power-law integral function, numerically computed as in \citet{RuedaBecerril:2017phd}. Finally, from equations~\eqref{eq:Le1} and~\eqref{eq:Le2} we get that
\begin{equation}\label{eq:Q0}
  Q_{0} = \dfrac{2 f_{\rm rec} L_{\rm B}}{3 \beta \Gamma^{2} V' m_{\rm e} c^{2} \gamma_{\min}'^{2 - p} \mathcal{P}(\gamma_{\max}' / \gamma_{\min}', p - 1)}.
\end{equation}

In the reconnection region we have that the magnetic energy available per electron in an electron-proton jet is $\sim \sigma m_{\rm p} c^2$. As we have mentioned, after reconnection takes place, a fraction of this energy $f_{\rm rec}$ goes into accelerated electrons. This fraction is model dependent as has been shown in \citet{Sironi:2015pe}. Additionally, the average energy per injected electron is $f_{\rm rec} \sigma m_{\rm p} c^2$, which means that the average Lorentz factor of the injected electron is \citep[e.g.,][]{Petropoulou:2016gi}
\begin{equation}
  \langle \gamma \rangle \sim f_{\rm rec} \sigma \dfrac{m_{\rm p}}{m_{\rm e}}.
\end{equation}

\subsubsection{Extrema of the non-thermal particles}
\label{sec:gmin}

From the average energy and average Lorentz factor of the injected electrons one finds that
\begin{equation}\label{eq:gmin-frec}
  \gamma_{\min}' = f_{\rm rec} \sigma \dfrac{m_{\rm p}}{m_{\rm e}} \left(\dfrac{p - 2}{p - 1}\right).
\end{equation}
The above result holds for $p > 2$ and $\gamma_{\max}' \gg \gamma_{\min}'$. On the other hand, if the distribution has a power-law index of $1 < p < 2$ we can make use of the result found in \citet{Sironi:2014sp}. In that work it was estimated that the mean energy per particle cannot exceed $(\sigma + 1) m_{\rm p} c^{2}$. From this it is deduced that the maximum Lorentz factor is given by
\begin{equation}\label{eq:gmax-sigma}
  \gamma_{\max}' = {\left(f_{\rm rec}(\sigma + 1) \dfrac{m_{\rm p}}{m_{\rm e}} \dfrac{2 - p}{p - 1}\right)}^{1 / (2 - p)}{\gamma_{\min}'}^\frac{1 - p}{2 - p}.
\end{equation}

The minimum and maximum Lorentz factors, $\gamma_{\min}'$ and $\gamma_{\max}'$, are set separately for high and low magnetized models. Regarding the value of $\gamma_{\max}'$ for the cases with low magnetization, i.e., with $p > 2$, is estimated by equating the acceleration rate of the electrons to the synchrotron cooling rate \citep{Dermer:2009}, i.e.,
\begin{equation}\label{eq:gmax-eacc}
  \gamma_{\max}' = {\left( \dfrac{6 \pi {\rm e}}{ \epsilon_{\rm acc} \sigma_{\rm T} B'} \right)}^{1/2},
\end{equation}
where the parameter $\epsilon_{\rm acc}$ could be interpreted as the number of gyrations the electron experience before it is injected into the system as part of the non-thermal distribution.

\subsection{Particle evolution}
\label{sec:part-distrib}

We will consider a one-zone model in which the emission region is a spherical blob of radius $R_{\rm b}'$ (see Eq.~\eqref{eq:blob-radius}) which moves with constant bulk Lorentz factor $\Gamma$ for a dynamical time. We assume that the accelerated particles radiate isotropically in this region. We perform our simulations using the numerical code \texttt{Paramo} \citep{RuedaBecerril:2020kn}. This code solves the Fokker-Planck equation using a robust implicit method \citep[see][]{Chang:1970co,Park:1996pe}, and for each time-step of the simulation the synchrotron, synchrotron self-absorption and inverse Compton emission (both synchrotron self-Compton, SSC, and external Compton, EIC) are computed with sophisticated numerical techniques \citep{Mimica:2012al,RuedaBecerril:2017mi,RuedaBecerril:2017phd}.

For the present work we will focus on solving the Fokker-Planck equation without diffusion terms, i.e.,
\begin{equation}\label{eq:FP}
  \dfrac{\upartial n'(\gamma', t')}{\upartial t'} + \dfrac{\upartial}{\upartial\gamma'} \left[ \dot{\gamma}'(\gamma', t') n'(\gamma', t') \right] = Q(\gamma', t') - \dfrac{n'(\gamma', t')}{t_{\rm esc}},
\end{equation}
where $n'$ is the electrons energy distribution (EED) in the flow comoving frame, $Q$ is a source term (see Eq.~\eqref{eq:inj-term}), and $t_{\rm esc} = t_{\rm dyn}$ is the electrons the average escape time. The electrons radiative energy losses are accounted for with the coefficient \citep{Rybicki:1979}:
\begin{equation}\label{eq:ener-loss}
  -\dot{\gamma}' = \dfrac{4 c \sigma_{\rm T}}{3 m_{\rm e} c^2} \beta_{\rm e}'^2 \gamma'^2 (u_{\rm B}' + u_{\rm BLR}'),
\end{equation}
where $\beta'_{\rm e}$ is the speed of the electron, in units of $c$, in the comoving frame.

\section{Results}
\label{sec:results}

\begin{table}
  \centering
  \begin{tabular}{cc}
    Parameter & Value \\
    \hline
    \hline
    $\theta_{\rm obs}$ & $2\degr$ \\
    $M_{\rm bh}$       & $10^{9} M_{\odot}$ \\
    $\eta_{\rm j}$     & 0.9 \\
    $\eta_{\rm d}$     & 0.1 \\
    $\eta_{\rm BLR}$   & 0.1 \\
    $\nu_{\rm BLR}$    & 2 eV / $h$ \\
    $f_{\rm rec}$      & 0.15 \\
    $s$                & 3.0 \\
    $\Gamma_0$         & 40 \\
    $\mu$              & 50, 70, 90 \\
    ($\sigma, p$)      & (1, 3.0), (3, 2.5), (10, 2.2), (15, 1.5), (20, 1.2)
  \end{tabular}
  \caption{Parameters of the present model. See text for a description of each of them.}
  \label{tab:tab1}
\end{table}

\begin{figure*}
  \includegraphics[width=\textwidth]{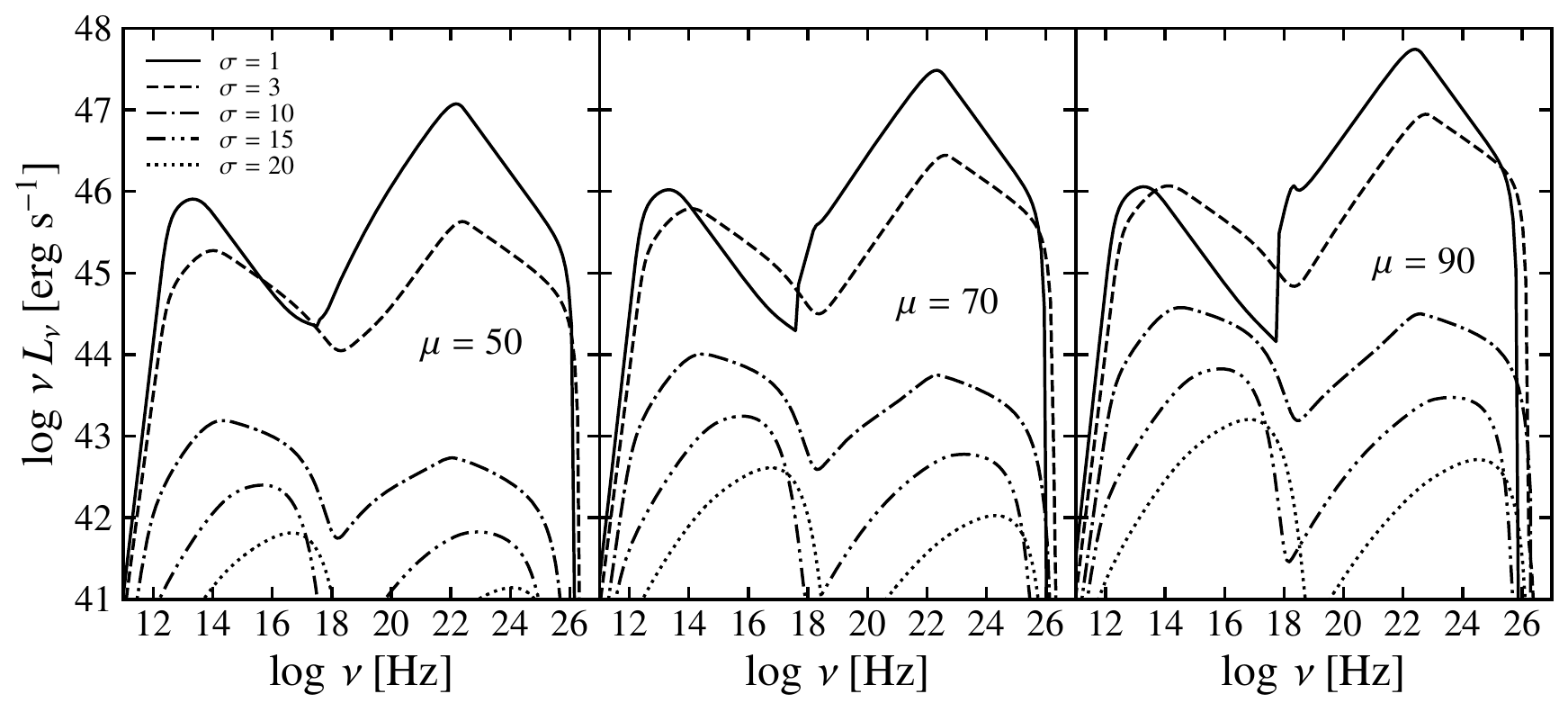}
  \caption{Sequence of blazar SEDs for varying model parameters. From left to right, each panel shows the averaged SEDs for different baryon loading $\mu = 50, 70, 90$, respectively. Solid, dashed, dot-dashed, dot-dot-dashed and dotted lines correspond to those simulations with $\sigma = 1, 3, 10, 15, 20$, respectively. The SEDs were averaged over 1~day after particles start being injected in the emission region.}
  \label{fig:BlazSeq}
\end{figure*}

In this section, we describe the results obtained from our simulations for different values of the parameters of the model. In our model, described in the previous section, we accomplished to reduce parameter space. In Tab.~\ref{tab:tab1} we summarize the parameters and values employed in the present work. As discussed below, the value of most of these parameters is constrained by either observations or theory.

The accretion disk and jet are parametrized by the black hole mass $M_{\rm bh}$, the radiative efficiency of the accretion disk $\eta_{\rm d}$, and the jet production efficiency $\eta_{\rm j}$. The values for these parameters were motivated by observations, theory and simulations. For instance, measurements of \citet{Bian:2003zh} and \citet{Davis:2011la} agree that, for quasars, $\eta_{\rm d} \sim 0.1$. Meanwhile, simulations by \citet{Tchekhovskoy:2012mc} show that $\eta_{\rm j}$ may vary between 0.3 and 0.9, depending on the spin of the black hole. Nevertheless, we studied the effect of changing $\eta_{\rm j}$ in our simulations. We observed that this parameter controls the luminosity of the synchrotron peak and, to a lesser extent, the luminosity of the EIC peak. With $\eta_{\rm j} = 0.9$ the bumps increase slightly, while for $\eta_{\rm j} = 0.3$ the objects are less luminosity, keeping qualitatively the same spectral features. The BLR is modeled by the covering factor $\eta_{\rm BLR}$, and the frequency of the external radiation field $\nu_{\rm BLR}$. Following the formulation by \citet{Ghisellini:2008ta}, we set $\eta_{\rm BLR} = 0.1$, while  $h \nu_{\rm BLR} = 2$~eV, which is an arbitrary value chosen between the characteristic hydrogen ionization frequencies H$\alpha$ and Ly-$\alpha$.

\begin{figure*}
  \includegraphics[width=\textwidth]{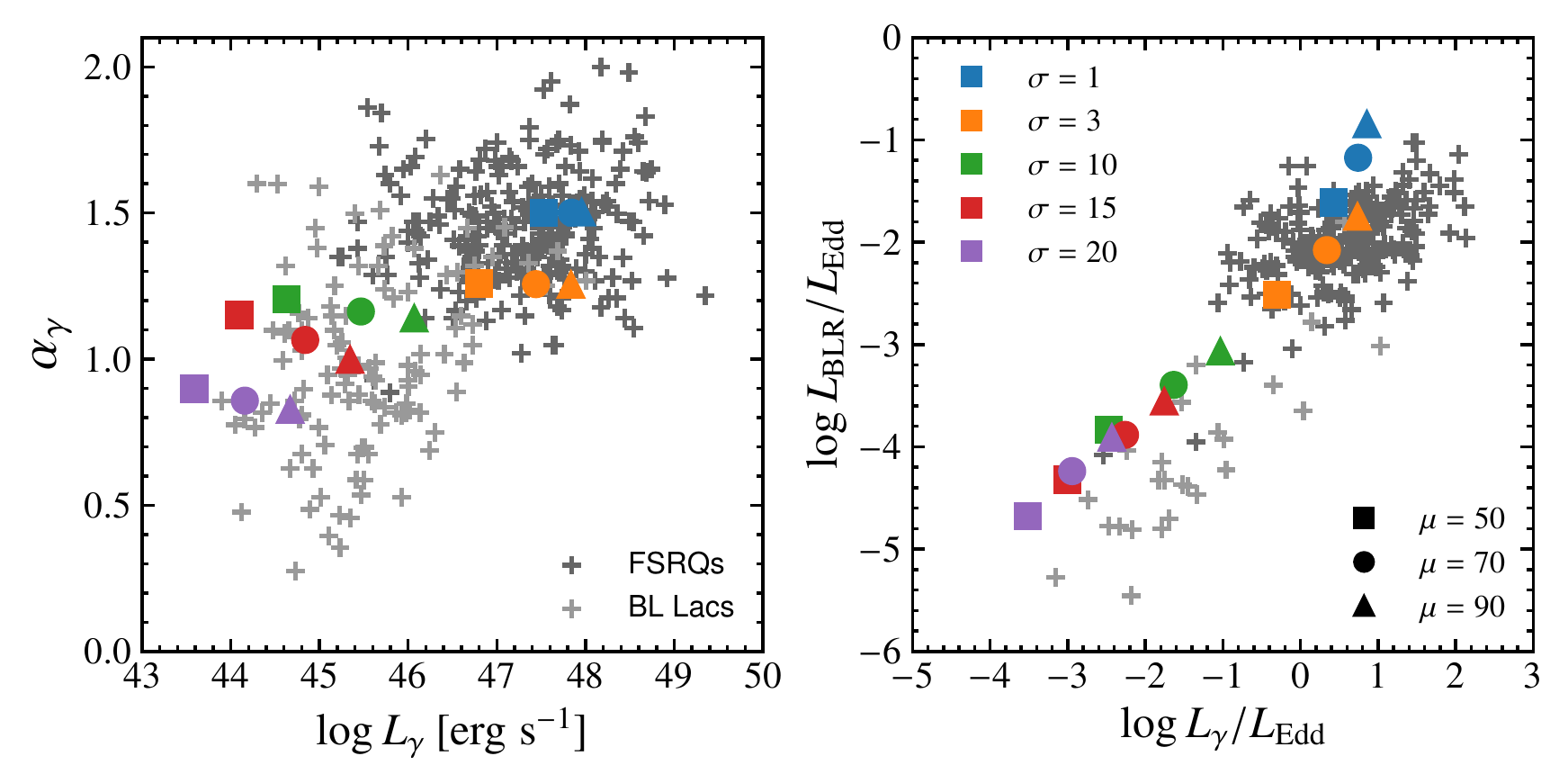}
  \caption{$\gamma$-ray spectral index $\alpha_{\gamma}$, $\gamma$-ray luminosity $L_{\gamma}$, and BLR luminosity $L_{\rm BLR}$. Observational data from \citet[][left panel]{Ghisellini:2011ta}, and \citet[][right panel]{Sbarrato:2014pa} is shown as dark and light gray crosses. Squares, circles and triangles depict the models with baryon loading $\mu = 50, 70, 90$, respectively. Blue, orange, green, red and purple colors show the simulation results with magnetization $\sigma = 1, 3, 10, 15$ and 20, respectively. {\it Left panel}: $\gamma$-ray energy spectral index $\alpha_{\gamma}$ as a function of the $\gamma$-ray luminosity $L_{\gamma}$. {\it Right panel}: Luminosity of the BLR $L_{\rm BLR}$ as a function of $L_{\gamma}$, both in units of the Eddington luminosity $L_{\rm Edd}$.}
  \label{fig:specidx}
\end{figure*}

The magnetic reconnection dissipation factor $f_{\rm rec}$ has been set to 0.15, following \citet{Petropoulou:2019si}. The power-law index $p$ of the injected particles, Eq.~\eqref{eq:inj-term}, has been estimated by \citet{Sironi:2015pe}, and more recently by \citet{Petropoulou:2019si}. Those works report that highly magnetized flows ($\sigma \gtrsim 10$) accelerate electrons with power-law indices in the range $1 \lesssim p \lesssim 2$, while mildly magnetized models ($\sigma \lesssim 10$) show electrons distributions with $p \gtrsim 2$. A highly magnetized jet will be associated with BL Lac objects, whereas the mildly magnetized to FSRQ jets. Finally, the extrema of the injected particle distribution, $\gamma_{\min}'$ and $\gamma_{\max}'$, for FSRQs are given by equations \eqref{eq:gmin-frec} and~\eqref{eq:gmax-eacc}, respectively, assuming that the most energetic electrons undergo $\approx 10^{6}$ gyrations before they are injected into the system (the exact choice for this parameter does not have an important effect on the results as long as $\gamma'_{\max} \gg \gamma'_{\min}$). Meanwhile, we know that the synchrotron peak of BL Lac-like simulations is given by $\gamma_{\max}'$, which is calculated using Eq.~\eqref{eq:gmax-sigma}. If we take a small value of $\gamma_{\min}'$, the synchrotron peak will shift to larger frequencies, some of them unrealistic, and not shown here. Using the synchrotron peak from radio observations as a guide, it is therefore possible to constrain $\gamma_{\min}'$ to a reasonable value of $\sim 1000$ for the injected distribution of particles in BL Lacs-like models.

Radio observations have shown that the bulk Lorentz factor of blazar jets ranges from a few to no more than 40 \citep[e.g.,][found that sources with $\Gamma > 40$ are extremely rare]{Lister:2016mo}. Assuming that blazar jets are ejected with similar baryon loading, a jet with $\mu$ of a few would imply that the jet will not be able to reach high magnetization and its Lorentz factor will be of order unity. Therefore, we estimate that a jet consistent with observations and simulations should have a baryon loading $\mu > 50$.

As we have mentioned in the previous section, our model resides on the hypothesis that all blazars are launched with similar baryon loading. In Fig.~\ref{fig:BlazSeq} we show the sequence of SEDs for three different values of $\mu$. The solid, dashed, dot-dashed, dot-dot-dashed and dotted lines correspond to magnetization $\sigma = 1, 3, 10, 15, 20$, respectively. FSRQs are the brightest of all blazars in all frequencies, their inverse Compton (IC) component tends to be louder than the synchrotron one, and $\nu_{\rm syn}$ falls in the infra-red. These features also appear in our simulations with the lowest magnetization, which we assumed as FSRQ-like. On the other hand, the main features observed in SEDs of BL Lac objects are a quieter IC component, $\nu_{\rm syn}$ in the UV--X-rays, and a harder spectral index in the $\gamma$-rays. We find that this is also the case for the highly magnetized cases. Finally, by contrasting all frames in Fig.~\ref{fig:BlazSeq} we can see the \emph{blazar sequence} trend \citep[cf. ][Fig.~12]{Fossati:1998ma} is favored for $\mu > 50$. The jets with larger baryon loading correspond to those sources with larger bulk Lorentz factor. From Eq.~\eqref{eq:dotm-gamma}, these sources correspond to the most efficient accretion disks which in turn correspond to those with most powerful jets (see Eq.~\eqref{eq:Mdot-Medd-rat}). This effect is more evident for the highly magnetized cases, whose luminosity increases for almost two orders of matnitude.

In Fig.~\ref{fig:specidx} we present our simulations with $\sigma = 1, 3, 10, 15, 20$ in blue, orange, green, red and purple points, respectively. Those simulations with baryon loading $\mu = 50, 70, 90$ are depicted in squares, circles and triangles, respectively. Observation data from \cite{Ghisellini:2011ta} is seen in light and dark gray crosses. On the left panel, we show the spectral index $\alpha_{\gamma}$ as a function of the bolometric luminosity $L_{\gamma}$ in the band 0.1-10~GeV \citep[cf. Fig.~1 in][]{Ghisellini:2011ta}. Observations here are presented in the 1LAC catalogue and range from $\gamma$-ray luminosity of 0.1 to 10 GeV and have known redshift. \citet{Ghisellini:2011ta} note that the division between BL Lacs and FSRQs is usually around $10^{46}$~ergs~s$^{-1}$, interpreted as a shift from an efficient accretion disk to a relatively inefficient disk. Our simulations show a similar trend: efficiently accreting sources with powerful jets (FSRQ-like) inhabit the area with $L_{\gamma} \gtrsim 10^{46}$~erg~s$^{-1}$ and softer $\gamma$-rays spectral index. Mild and highly magnetized simulations fall in the area of BL Lac objects with low $\gamma$-rays luminosity.

On the right panel of Fig.~\ref{fig:specidx} we show the BLR luminosity, $L_{\rm BLR}$, as a function of $L_{\gamma}$, both in units of the Eddington luminosity $L_{\rm Edd}$, together with observational data points from  Fig.~1, right panel, in \citet[][]{Sbarrato:2014pa}. According to this paper, those sources with a stronger emission lines, i.e., showing a more luminous BLR, appear louder in the $\gamma$-ray band. The latter being FSQRs. In our simulations, the corresponding ones with a more luminous BLR are those with larger $\Gamma$. Our model states that these objects have larger Eddington ratio (see Eq.~\eqref{eq:dotm-gamma}), i.e., that would correspond to highly efficient accretion objects.

\begin{figure*}
  \includegraphics[width=\textwidth]{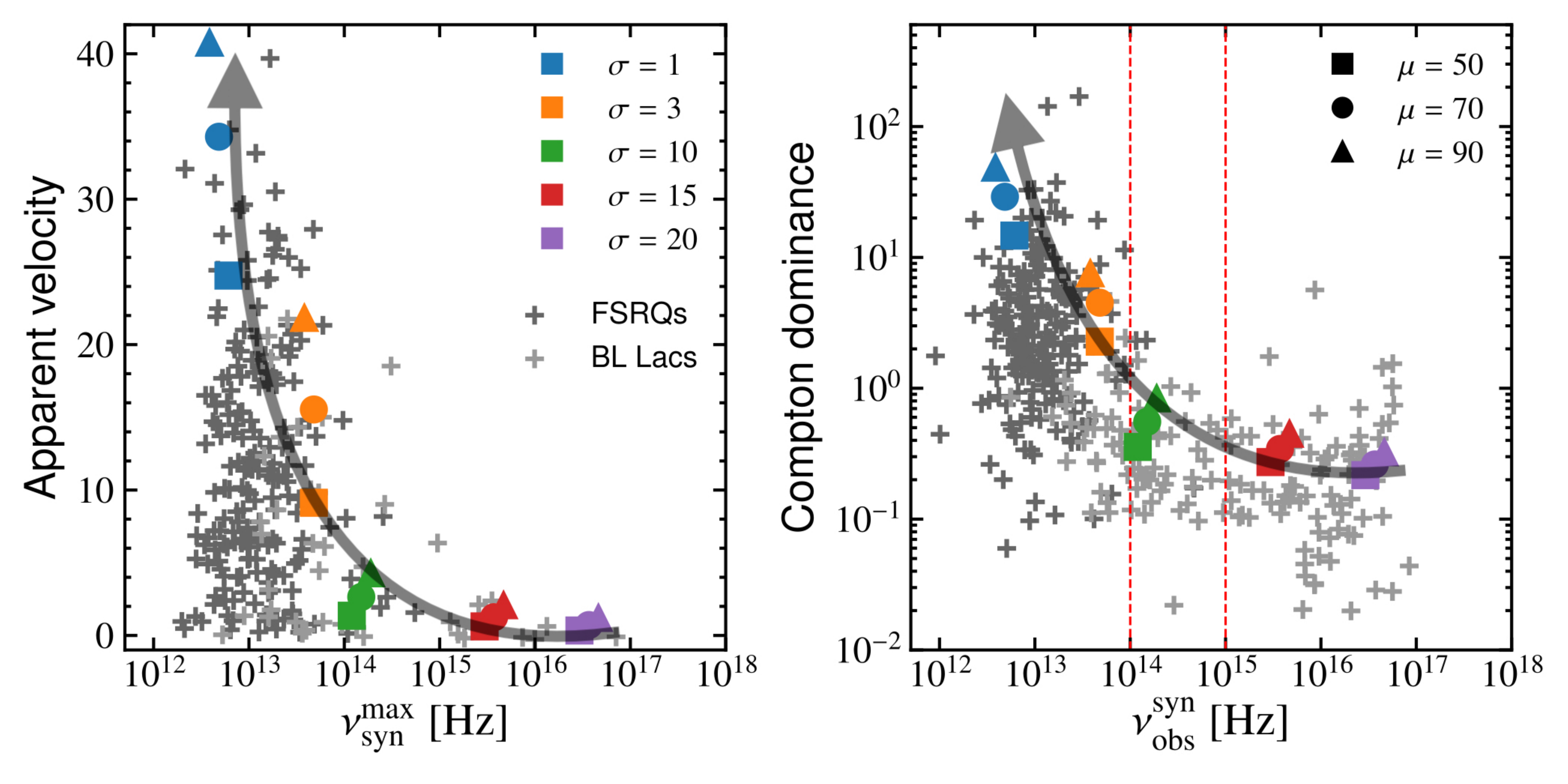}
  \caption{Apparent velocity, Compton dominance and synchrotron peak. Similar to Fig.~\ref{fig:specidx}, squares, circles and triangles depict the models with baryon loading $\mu = 50, 70, 90$, respectively. Blue, orange, green, red and purple colors show the simulation results with magnetization $\sigma = 3, 10, 20, 50$, respectively. The gray transparent arrow shows the increasing trend of the jet luminosity, $L_{\rm j}$. {\it Left panel}: We show the apparent velocity as a function of the synchrotron peak frequency $\nu_{\rm syn}$. Observational data from \citet{Lister:2019ho}. {\it Right panel}: We show the Compton dominance as a function of the synchrotron peak. In red dashed vertical lines we separate the LBL ($\lesssim 10^14$~Hz), IBL ($\gtrsim 10^{14}$~Hz and $\lesssim 10^{15}$~Hz) and HBL ($\gtrsim 10^{15}$~Hz) regions. Observational data from \citet{Finke:2013co}.}
  \label{fig:G-vsyn}
\end{figure*}

In the same manner, in Fig.~\ref{fig:G-vsyn} we present our simulations with $\sigma = 1, 3, 10, 15, 20$ in blue, orange, green, red and purple points, respectively. Baryon loadings $\mu = 50, 70, 90$ are shown in squares, circles and triangles, respectively. Light and dark gray crosses correspond to BL Lacs and FSRQs sources, respectively. On the left panel we show the apparent velocity%
\footnote{The apparent velocity is calculated according to the following expression:
  \[
    v_{\rm app} = \dfrac{v \sin\theta_{\rm obs}}{1 - \frac{v}{c} \sin\theta_{\rm obs}},
  \]
  where $v = \beta c$ is the bulk speed of the flow.
} of our synthetic objects. The observational data correspond to the data in the MOJAVE survey, reported in \citet{Lister:2019ho}. A translucent gray arrow draws the trend of increment of the jet luminosity. In this plot we can appreciate how the synchrotron peak $\nu_{\rm syn}$ of our simulations is similar for each magnetization. The apparent velocity is bulk Lorentz factor dependent due to relativistic boosting. This effect is clear for those objects with larger $\Gamma$ (blue and orange points), which correspond to those simulations with more powerful jets. Our simulations with powerful jets concur with FSRQs as assumed. This is the case as well with highly magnetized objects. These objects represent the less powerful jets, and fall well in the region of BL Lacs.

In the leptonic model of blazars, the \emph{Compton dominance} is defined as the ratio of luminosities between the IC and the synchrotron components of their SED. On the right panel of Fig.~\ref{fig:G-vsyn} we contrast the Compton dominance and $\nu_{\rm syn}$ of our synthetic sources with the observational data reported in \citet{Finke:2013co}, depicted as gray crosses. These sources are presented in the 2LAC clean sample where all had known redshift and could clearly be classified. In that same work, sources with unknown redshift were also taken into account, finding that the relation between Compton dominance and synchrotron peak frequency have a physical origin rather than it being a redshift selection effect. Regarding our simulations, we can observe that all our simulations fall within the observational points. The gray transparent arrow shows the trend of increment of the jet luminosity. Our simulations show that, keeping $\mu$ constant, changing the magnetization will give the transition from synchrotron-dominant (highly magnetized) to Compton-dominant and $\gamma$-ray loud sources.


\section{Discussion}
\label{sec:discuss}

\begin{figure}
  \includegraphics[width=\columnwidth]{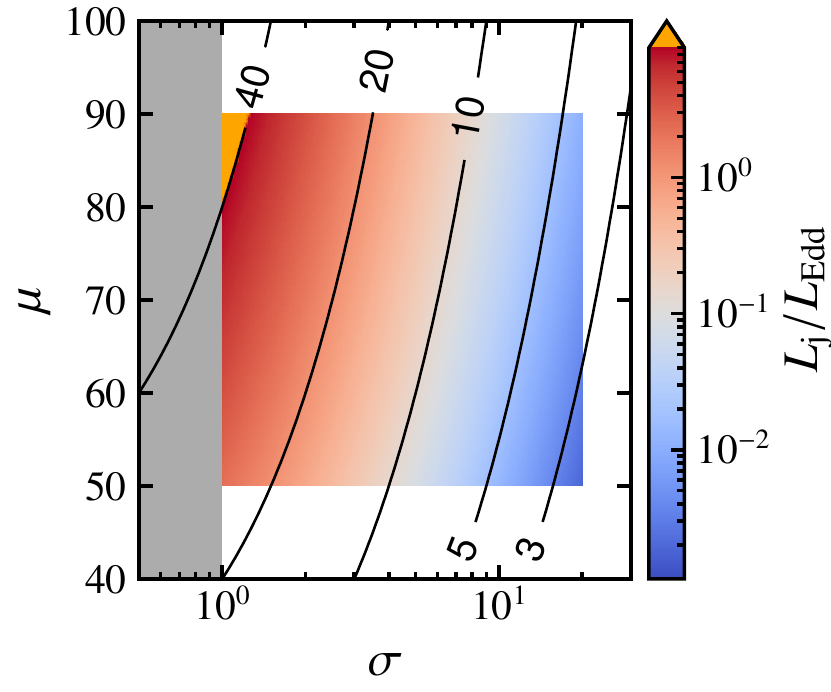}
  \caption{The baryon loading $\mu$ as a function of magnetization $\sigma$. Contour lines correspond to the bulk Lorentz factor (see Eq.~\eqref{eq:mu}). The color gradient shows the jet luminosity $L_{\rm j}$ (see Eq.~\eqref{eq:Mdot-Medd-rat}), and the gray area depicts the $\sigma<1$ region. (The vertical axis has been changed accordingly. The colored area was adjusted to the range of values studied in this work.)}
\label{fig:musigam}
\end{figure}

According to our model, BL Lacs are those blazars with largest magnetization ($\sigma\gtrsim 10$) at the dissipation region. FSRQs, on the other hand, are those with powerful jets but with low/mild magnetization ($\sigma\lesssim 10$) at the blazar zone. In Fig.~\ref{fig:musigam}, it is shown the relation between the main parameters of our study: the magnetization $\sigma$, the bulk Lorentz factor $\Gamma$, and the baryon loading $\mu$, as prescribed by the $\mu\sigma\Gamma$ relation~\eqref{eq:mu}. In color gradient we have included the corresponding jet luminosity $L_{\rm j}$, in units of the Eddington luminosity $L_{\rm Edd}$ (see Eq.~\eqref{eq:Mdot-Medd-rat}). The $\mu\sigma\Gamma$ relation constrains these objects to have a mild baryon loading since our model stands on the assumption that blazars are launched with similar baryon loading. Jets launched with $\mu > 90$ would give values of $\Gamma$ beyond those inferred from radio observations, for those cases with low magnetization. If blazars, on the other hand, were launched with low baryon loading, e.g., $< 50$, the resulting $\Gamma \sim 1$ for the highly magnetized cases would contradict both simulations and observations. These scenarios have been discarded from our analysis. BL Lac objects, as blazars with low jet luminosity, fall in the blue--gray region with $\lesssim 10^{-1} L_{\rm Edd}$. According to our results (described in Sec.~\ref{sec:results}), this same region corresponds to our simulations with high magnetization. FSRQs, the most powerful of observed blazars, fall in the the gray--red region. Jets with super-Eddington power, i.e., those cases with $\dot{m} \sim 1$, belong to the orange region in upper-left corner (see App.~\ref{sec:acc-rate}).

Mildly magnetized blazars, e.g., $\sigma = 10$, develop a particular behavior. These models have an Eddington rate $L_{\rm j} / L_{\rm Edd} \sim 0.1$, synchrotron peak $\nu_{\rm syn} \gtrsim 10^{14}$~Hz, like some FSRQs. However, their IC component is less ($\mu = 50$) or similar ($\mu = 90$) in luminosity to the synchrotron component, and the $\gamma$-ray spectral index is harder; characteristics of BL Lac objects. According to \citet{Padovani:2019oi}, the object TXS 0506+056, a ``masquerading'' BL Lac object, shows properties like $10^{46} \lesssim L_{\gamma} / (\mathrm{erg~s^{-1}}) \lesssim 10^{48}$ and $10^{14} \lesssim \nu_{\rm syn} / \mathrm{Hz} \lesssim 10^{15}$. According to our simulations, mildly magnetized ones (dot-dashed lines in Fig.~\ref{fig:BlazSeq}, and green dots in Figures~\ref{fig:specidx} and~\ref{fig:G-vsyn}) have also these features. Moreover, in Fig.~\ref{fig:musigam} we can place our mildly magnetized model in the region $L_{\rm j} / L_{\rm Edd} \approx 0.1$, which would correspond to an Eddington ratio $\dot{m} \gtrsim 0.01$.

In this work we have associated the most extreme accretion systems with blazar jets with large $\Gamma$. The bright accretion disk may dominate the ionizing flux received by the gas clouds living in the BLR, obscuring the central engine and populating that space with a denser photon field from the reprocessed disk radiation. A denser photon field, in conjuction with the larger blulk $\Gamma$, translate into a more luminous EIC component of the blazar SED. A denser external radiation field would also mean a strong cooling factor $\dot{\gamma}$, steepening the EED. This agrees with recent findings by \citet{Keenan:2020me}. They agree with the scenario in which powerful blazar have a broad-emitting gas surrounding the core. This also agrees with recent findings of \citet{Zhang:2020zh}, regarding the jet properties of other kind of $\gamma$-ray emitting AGNs known as Compact Steep-spectrum Sources.

Regarding the core surrounding environment, according to \citet{Ghisellini:2011ta} and \citet{Sbarrato:2014pa}, there is a clear division between FSRQs and BL Lacs in the $L_{\rm BLR}$--$L_{\gamma}$ plane at $L_{\rm BLR} / L_{\rm Edd} = 5 \times 10^{-4}$. According to our model, this divide is not so clear. As we have mentioned before, mildly magnetized simulations have been setup as FSRQ-like, however, comparing with observables, these show BL Lac features as well. It may be the case that there is not such a sharp divide between BL Lacs and FSRQs.

Looking back into the Compton dominance plot (right panel of Fig.~\ref{fig:G-vsyn}), if we focus on a particular value of $\mu$, e.g., triangles, one can move through all the observational region by increasing the jet luminosity, following the gray translucent arrow. In other words, blazar jets may indeed launch with similar baryon loading. In low $\dot{m}$ systems, a fainter accretion disc means low density external photon field surrounding the emission region and a less powerful jet (blue region in the lower right region of Fig.~\ref{fig:musigam}). Jets in low $\dot{m}$ systems have mostly the synchrotron photons produced \emph{in situ} as seed photons for upscattering, showing a dim EIC, just like BL Lac objects whose inner core shows no significant sign of a broad-emitting gas. The SSC component is therefore dominant in these sources, although not expected to be as $\gamma$-ray loud as the EIC of powerful jets, where Doppler boosting plays a leading role in enhancing the EIC component. 

In both panels of Fig.~\ref{fig:G-vsyn}, the synchrotron peak $\nu_{\rm syn}$ of powerful jet simulations corresponds to the synchrotron frequency of the cooling break of the EED%
\footnote{The cooling break of a particle energy distribution corresponds to the energy at which the distribution changes slope and is given by the cooling factor $\dot{\gamma}$ in the kinetic equation~\eqref{eq:FP}. This point depends on how fast particles are being cooled down. The analysis of the cooling stages of the EED in the emission region of our blazar model here presented is beyond the scope of this work.%
}. It turns out that, because of strong radiative losses, powerful jets have a synchrotron peak deeper into the far infrared (bellow these frequencies, most of the synchrotron emission is self-absorbed). On the oposite side, $\nu_{\rm syn}$ of our simulated BL Lac objects (i.e., simulations with high magnetization and $1 < p < 2$), corresponds to the synchrotron frequency of the maximum Lorentz factor of the EED, $\gamma_{\max}'$, given by Eq.~\eqref{eq:gmax-sigma}. For these cases, in contrast with simulations with low magnetization, $\gamma_{\max}'$ is highly dependent on $\gamma_{\min}'$. This setup of the EEDs in our model doesn't give any restriction or upper limit for the synchrotron peak $\nu_{\rm syn}$ \citep[see][]{Keenan:2020me}. However, from Eq.~\eqref{eq:gmax-sigma}, observations can indeed constrain the value of $\gamma_{\rm min}'$.

\section{Conclusions}
\label{sec:conclusions}

In the present work, we have applied a simple idea that accounts for the blazar sequence and several observable features of the blazar population. This model relies on the idea that all jets are launched with similar energy per baryon, independently of their power. FSRQs, those with the most powerful jets, manage to accelerate to high bulk Lorentz factor and have luminosities $\gtrsim 0.2 L_{\rm Edd}$. FSRQ-like simulations were set to have a rather modest magnetization in the emission region and a steep particle energy distribution. Our predicted SEDs of these models show similar features as actual FSRQs observations: peak synchrotron $\nu_{\rm syn} \lesssim 10^{14}$~Hz, Compton dominance, soft spectra in the $\gamma$-rays, and are $\gamma$-ray louder. In the case of BL Lacs, the jet does not achieve a very high bulk Lorentz factor, leading to more magnetic energy available for non-thermal particle acceleration. According to our model (see Sec.~\ref{sec:model}), these sources develop high synchrotron peak, weaker Compton component, and harder emission spectra at frequencies $\gtrsim$ GeV.

With our model and simulations reported in this work, we were able to recover observables of blazars. Namely, the \emph{blazar sequence} was (qualitatively) reproduced, in a similar manner as it was first reported by \citet{Fossati:1998ma}, for those models with mild baryon loading. This result constrains the energy per baryon of blazar jets to $50 \lesssim \mu \lesssim 80$. The L-like region observed for the apparent velocity and Compton dominance as functions of $\nu_{\rm syn}$ was also recovered by changing $L_{\rm j}$, assuming that it tracks $\dot{m}$. With our simple model we are also able to show that the brightness of the BLR scales linearly with the $\gamma$-rays loudness of the source.

Finally, we propose an indirect method to estimate $\gamma_{\min}'$ for BL Lacs. From the value of $\nu_{\rm syn}$ given by observations we can directly calculate $\gamma_{\max}'$. Following Eq.~\eqref{eq:gmax-sigma} we are therefore able to calculate $\gamma_{\min}'$. PIC simulations of magnetic reconnection may be able to test whether our adopted values are reasonable.

It is worth highlighting the particular case in which an FSRQ-like simulation (green points in figures~\ref{fig:specidx} and~\ref{fig:G-vsyn}), is in fact $\gamma$-ray quieter. This object would in principle have a mild Eddington rate $\dot{m}$, and a mildly luminous BLR. However it is not powerful enough to develop an IC component louder than its synchrotron component. Additionally, it has a harder spectral index $\alpha_{\gamma}$, and emits close the TeV band, just like BL Lacs. Similar ``contradicting'' properties have also been observed in objects like TXS 0506+056.

In summary, our model assumes that all jets are injected with energy per baryon in a narrow range $50 \lesssim \mu \lesssim 80$ and that the jet bulk Lorentz factor and power scale positively with the accretion rate, and can account for or predict:
\begin{itemize}
  \item That $\dot{m}$ controls many of the observable features of blazars such as the high-energy spectral index and luminosity, the brightness of the BLR, the apparent speed, and the synchrotron spectrum and synchrotron peak frequency.
  \item Sources that are $\gamma$-ray brighter have softer $\gamma$-ray spectral index $\alpha_{\gamma}$. Lower values of $\alpha_{\gamma}$ (i.e., harder spectra) were found for the $\gamma$-ray quieter sources.
  \item The BLR luminosity $L_{\rm BLR}$ scales linearly with the $\gamma$-ray luminosity of the object.
  \item Fastest objects have low-frequency synchrotron peak $\nu_{\rm syn}$ while objects with intermediate-to-high synchrotron peak move rather slow.
  \item Low jet luminosity sources are non-Compton dominant but high synchrotron-peaked, whereas those with higher Compton dominance have a $\nu_{\rm syn} \lesssim 10^{13}$~Hz.
\end{itemize}

\section*{Acknowledgements}

It is a pleasure to thank Matt Lister for useful comments, and the anonymous referee for insightful comments and valuable suggestions. The research was partly supported by Fermi Cycle 12 Guest Investigator Program \#121077. JMRB acknowledges the support from the Mexican National Council of Science and Technology (CONACYT) with the Postdoctoral Fellowship under the program Postdoctoral Stays Abroad. DG acknowledges support from the NASA ATP NNX17AG21G, the NSF AST-1910451 and the NSF AST-1816136 grants. This research was supported in part through computational resources provided by Information Technology at Purdue, West Lafayette, IN, USA.

\section*{Data availability}

Simulations were performed making use of the code \texttt{Paramo} \citep{RuedaBecerril:2020kn}. The version used for the present work is available under request to the corresponding author.

Source points (gray crosses) in Fig.~\ref{fig:specidx} were taken from \citet[][left panel]{Ghisellini:2011ta}, and \citet[][right panel]{Sbarrato:2014pa}. Source points (gray crosses) in Fig.~\ref{fig:G-vsyn}, left panel, were taken from \citet{Lister:2019ho}. Source points (gray crosses) in Fig.~\ref{fig:G-vsyn}, right panel, were taken from \citet{Finke:2013co}. WebPlotDigitizer was used for the data extraction.

\bibliographystyle{mnras}
\bibliography{blazMag} 

\appendix

\section{Accretion rate}
\label{sec:acc-rate}

\begin{figure*}
  \centering
  \includegraphics[width=\textwidth]{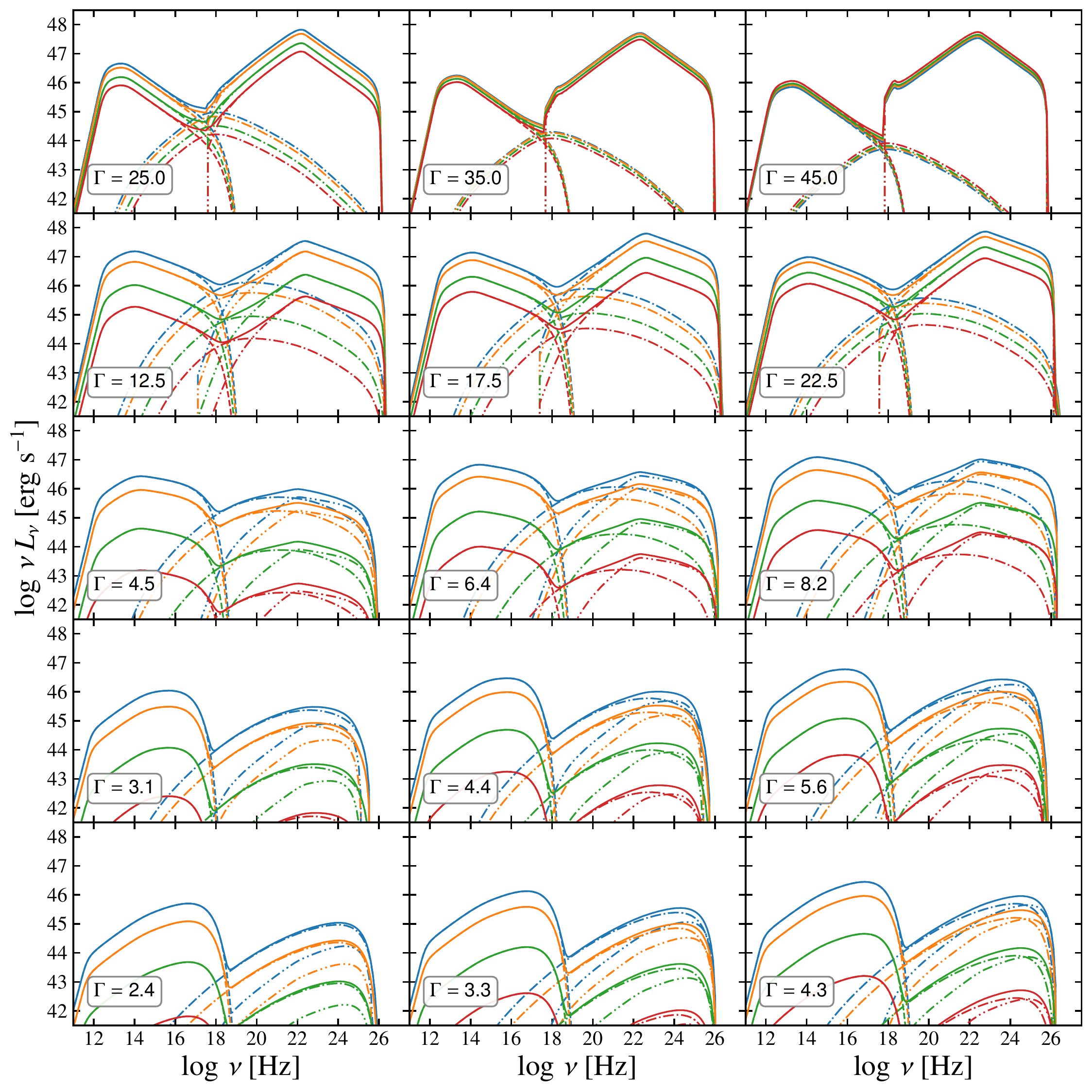}
  \caption{In this figure we show the averaged SEDs of our simulations with $\mu = 50, 70$ and 90 in the left, middle and right columns, respectively. Simulations with $(\sigma, p) = (1, 3.0), (3, 2.5), (10, 2.2), (15, 1.8)$, and $(20, 1.5)$ are shown from top to bottom, respectively. The solid, dashed, dot-dashed and dot-dot-dashed lines correspond to the total, synchrotron, SSC and EIC components, respectively. In blue, orange, green and red are depicted the simulations with accretion index $s = 1.5, 2.0, 3.0, 4.0$, respectively. The normalization bulk Lorentz factor in Eq.~\eqref{eq:dotm-gamma} is set to $\Gamma_0 = 40$. The spectra are averaged over 1~dy since particles start being injected into the emitting blob. The value of the bulk Lorentz factor $\Gamma$ shown in each panel is given by Eq.~\eqref{eq:mu}.}
  \label{fig:SEDs}
\end{figure*}

In the present section we will describe the parametrization of our model. In our formulation, the accretion rate parameter is given by Eq.~\eqref{eq:dotm-gamma}\footnote{A series of simulations with random values of $\Gamma$ and $\dot{m}$ were performed to rule out any overlooked relation. Obtaining, as expected, no apparent correlation between the two.}. The main effects of changing values of the accretion index $s$ are shown in Fig.~\ref{fig:SEDs}. There we show the averaged SEDs from simulations with $\mu = 50, 70, 90$ (columns from left to right, respectively), $(\sigma, p) = (1, 3.0), (3, 2.5), (10, 2.2), (15, 1.8)$, and $(20, 1.5)$ (rows from top to bottom, respectively), and $s = 1.5, 2.0, 3.0, 4.0$ (blue, orange, green and red lines, respectively). The synchrotron, SSC, EIC and total fluxes are depicted in dashed, dot-dashed, dot-dot-dashed and solid lines, respectively. For all simulations we set $\Gamma_{0} = 40$, and in each panel it is noted the corresponding bulk Lorentz factor, $\Gamma$, according to Eq.~\eqref{eq:mu}.

The first three rows (top to bottom) correspond to models setup FSRQ-like, i.e., with low-to-mild magnetization and $p > 2$. The first two are the brightest and the most Compton dominant. In fact, the EIC component is the dominant radiative process in all this set of simulations. Not so the middle-row ones, which show an EIC component with similar brightness, or dimmer, than the synchrotron component. BL Lac-like models are those with higher magnetization and lower $\Gamma$ (last two rows from top to bottom). These simulations show synchrotron, SSC and EIC components with similar luminosities.

The main effect that the normalization bulk Lorentz factor $\Gamma_{0}$ has on our simulations is the overall increase/decrease in luminosity. In the same manner, we noticed in the SEDs that by increasing the accretion index $s$, overall brightness decreases, but the overall spectral structure remains the same. Furthermore, this effect occurs regardless of the magnetization and baryon loading. From these results we can conclude that $\dot{m}$ regulates the intensity of the SEDs without changing any local nor broadband spectral feature. This was expected according to Eq.~\eqref{eq:Mdot-Medd-rat}, which tells us that $\dot{m}$ is a measure of $L_{\rm j}$. The cases that have reached the super-Eddington limit, i.e., those models with $\dot{m} \geq 1$, appear in uppermost right panel. In our setup, this frontier is set by the parameter $\Gamma_{0}$.

\bsp	
\label{lastpage}
\end{document}